\documentclass[prb,floats,twocolumn,showpacs,superscriptaddress]{revtex4}
\usepackage{amsmath}
\usepackage{graphicx}
\usepackage{dcolumn}
\usepackage{hyperref}
\newcommand{\smallwidth}{0.5\columnwidth}
\newcommand{\figwidth}{1\columnwidth}
\newcommand{\midwidth}{0.75\columnwidth}

\newcommand{\PP}{\mathcal{P}}

\newcommand{\tj}{$t{-}J$ }

\begin{document}
\title{Bond-order modulated staggered flux phase
for the $t{-}J$ model on the square lattice}

\author{C\'edric Weber}
\affiliation{
  Institut Romand de Recherche Num\'erique en Physique des Mat\'eriaux (IRRMA),
  PPH-Ecublens, CH-1015 Lausanne}
%%IRRMA, PPH-Ecublens, CH-1015 Lausanne, Switzerland}

\author{Didier Poilblanc}
\email{didier.poilblanc@irsamc.ups-tlse.fr}
\affiliation{ Laboratoire de Physique Th\'eorique UMR 5152, C.N.R.S. \&
Universit\'e de Toulouse, F-31062 Toulouse, France }
\affiliation{
Institute of Theoretical Physics,
Ecole Polytechnique F\'ed\'erale de Lausanne,
BSP 720,
CH-1015 Lausanne,
Switzerland}

\author{Sylvain Capponi}
\affiliation{ Laboratoire de Physique Th\'eorique UMR 5152, C.N.R.S. \&
Universit\'e Paul Sabatier, F-31062 Toulouse, France }

\author{Fr\'ed\'eric Mila}
\affiliation{
Institute of Theoretical Physics,
Ecole Polytechnique F\'ed\'erale de Lausanne,
BSP 720,
CH-1015 Lausanne,
Switzerland}

\author{Cyril Jaudet}
\affiliation{ Laboratoire de Physique Th\'eorique UMR 5152, C.N.R.S. \&
Universit\'e Paul Sabatier, F-31062 Toulouse, France }

\begin{abstract}
Motivated by the observation of inhomogeneous patterns in some
high-T$_c$ cuprate compounds, several variational
Gutzwiller-projected wave-functions with built-in charge and bond
order parameters are proposed for the extended $t-J-V$ model on
the square lattice at low doping. First, following a recent
Gutzwiller-projected mean-field approach by one of us
(Phys.~Rev.~B. {\bf 72}, 060508(R) (2005)), we investigate, as a
function of doping and Coulomb repulsion, the stability
of the staggered flux phase with respect to small spontaneous
modulations  of squared unit cells ranging
from $2\times 2$ to $\sqrt{32}\times\sqrt{32}$.
It is found that a $4\times 4$
bond-order (BO) modulation appears spontaneously on top of
the staggered flux pattern for hole doping around $1/8$.
A related wave-function is then constructed and
optimized accurately and its properties
studied extensively using an approximation-free variational Monte
Carlo scheme. Finally, the competition of
the BO-modulated staggered flux wave-function
w.r.t. the d-wave RVB wave-function or the commensurate flux state
is investigated. It is found that a short range Coulomb repulsion
penalizes the d-wave superconductor and
that a moderate Coulomb repulsion brings them very close in energy.
Our results are discussed in connection to the STM
observations in the under-doped regime of some cuprates.

\end{abstract}
\maketitle

\section{Introduction: models and methods}
\label{intro}

The observation of a d-wave superconducting gap in the high-$T_c$
cuprate superconductors suggests \cite{anderson_hgtc_hubbard} that
strong correlations are responsible for their unconventional
properties and superconducting behavior. The two-dimensional (2D)
\tj model is one of the simplest effective models proposed
\cite{Zhang-Rice} to describe the low energy physics of these
materials,
\begin{eqnarray}
  H_{t-J}&=& -t\sum\limits_{\langle i,j \rangle,\sigma }{
    \left( c_{i,\sigma }^{\dagger}c_{j,\sigma} + h.c. \right)} \nonumber \\
  &&+J\sum\limits_{\langle i,j \rangle}{ \mathbf{S}_{i}\cdot
      \mathbf{S}_{j} }
  \label{Htj}
\end{eqnarray}
The electrons are hopping between nearest neighbor sites of a
square lattice leading to a kinetic energy term (first term of
\ref{Htj}) as well as an exchange energy due to their spin
interaction (second term), where $\mathbf{S}_{i}$ denotes the spin
at site $i$:
$\mathbf{S}_{i}=\frac{1}{2}c_{i,\alpha}^{\dagger}\vec{\sigma}_{\alpha,\beta}
c_{i,\beta}$ and $\vec{\sigma}$ is the vector of Pauli matrices.
$\langle i,j \rangle$ stands for a pair of nearest neighbors.
$H_{t-J}$ operates only in the subspace where there are no doubly
occupied sites, which can be formally implemented by a Gutzwiller
projector (see later). In the following we set $\left\vert
t\right\vert =1$ (unless specified otherwise) and we adopt a
generic value of $t/J=3$ throughout the paper. Because of the
particle-hole symmetry in the square lattice the sign of $t$ does
not play any role. Although this model is formulated in a very
simple form, the nature of the quantum correlations makes its
physics very rich, and even the ground state of the \tj
hamiltonian was not yet characterized for finite doping and large
cluster size. However, the \tj model was investigated extensively
by unbiased numerical techniques \cite{review_Dagotto} as well as
by mean-field \cite{kotliar_fluxphases} and variational
Monte-Carlo approaches \cite{yokoyama_mcv_supra,gros_mcv_supra}.
All approaches found a d-wave superconducting phase and a phase
diagram which accounts for most of the experimental features of
the high-$T_c$ cuprates\cite{paramekanti_mcv,RVB2}. In the limit
of vanishing doping (half-filling), such a state can be viewed as
an (insulating) resonating valence bond (RVB) or \emph{spin
liquid} state. In fact, such a state can also be written (after a
simple gauge transformation) as a staggered flux state (SFP)
\cite{kotliar_fluxphases,half-flux}, i.e. can be mapped to a
problem of free fermions hopping on a square lattice thread by a
staggered magnetic field.

Upon finite doping, although such a degeneracy breaks down, the
SFP remains a competitive (non-superconducting) candidate with
respect to the d-wave RVB superconductor \cite{staggered_flux}. In
fact, it was proposed by P.A.~Lee and collaborators
\cite{PLee1,PLee2,PLee3} that such a state bears many of the
unconventional properties of the pseudo-gap \emph{normal} phase of
the cuprate superconductors. This simple mapping connecting a free
fermion problem on a square lattice under magnetic
field~\cite{Hofstadter} to a correlated wave-function (see later
for details) also enabled to construct more exotic flux states
(named as \emph{commensurate flux states}) where the fictitious
flux could be uniform and commensurate with the particle
density~\cite{Flux_phases_0,Flux_phases_1}. In this particular
case, the unit-cell of the tight-binding problem is directly
related to the rational value of the commensurate flux.

With an increasing number of materials and novel experimental
techniques of constantly improving resolution, novel features in
the global phase diagram of high-T$_c$ cuprate superconductors
have emerged. One of the most striking is the observation, in some
systems, of a form of local electronic ordering, especially around
$1/8$ hole doping. Indeed, recent scanning tunnelling
microscopy/spectroscopy (STM/STS) experiments of underdoped
Bi$_2$Sr$_2$CaCu$_2$O$_{8+\delta}$  (BSCO) in the pseudogap state
have shown evidence of {\it energy-independent} real-space
modulations of the low-energy density of states
(DOS)~\cite{STM-BSCO,STM-BSCO-note,STM_fourfold_structure,STM_fourfold_charge_order}
with a spatial period close to four lattice spacings. A similar
spatial variation of the electronic states has also been observed
in the pseudogap phase of Ca$_{2-x}$Na$_x$CuO$_2$Cl$_2$ single
crystals ($x=0.08\sim0.12$) by similar STM/STS
techniques~\cite{STM2}. Although it is not clear yet whether such
phenomena are either generic features or really happening in the
bulk of the system, they nevertheless raise important theoretical
questions about the stability of such structures in the framework
of our microscopic strongly correlated models.

In this paper, we analyze the stability and the properties
of new inhomogeneous phases
(which may compete in certain conditions with the d-wave superconducting RVB
state) by extending the previous mean-field and variational treatments of the
RVB theory.
In addition, we shall also
consider an extension of the simple \tj model, the
$t-J-V$ model, containing a Coulomb repulsion
term written as,
\begin{equation}
 V=\frac{1}{2}\sum\limits_{i \ne j}
V(|i-j|) \left( n_i-n \right) \left( n_j-n \right)\ ,
\end{equation}
where $n$ is the electron density ($N_e/N$, $N_e$ electrons on a
$N$-site cluster). Generically, we assume a screened Coulomb potential :
\begin{equation}
 V(r)=V_0\frac{\exp^{-r/\ell_0}}{r}\ ,
\end{equation}
where we will consider two typical values $\ell_0=2,4$ and
$V_0\in[0,5]$ and where the distance $r$ is defined (to minimize
finite size effects) as the \emph{periodized} distance on the
torus \cite{note_distance}.
The influence of this extra repulsive term in the competition between
the d-wave RVB state and some inhomogeneous phases is quite subtle
and will be discussed in the following.

To illustrate our future strategy, let us recall in more details
the simple basis of the RVB theory. It is based
on a mean-field hamiltonian which is of BCS type,
\begin{multline}
\label{eq:BCS}
  H_{\rm BCS}=\sum\limits_{\langle i,j \rangle,\sigma}
  \left( -\chi_0 c_{i\sigma }^{\dagger} c_{j\sigma} + h.c. \right) \\
  +\sum\limits_{\langle i,j \rangle}
    \left( \Delta_{i,j} c_{i\uparrow}^{\dagger}
    c_{j\downarrow}^{\dagger}
    + h.c.\right) -\mu\sum\limits_{i,\sigma } n_{i,\sigma }\ ,
\end{multline}
where $\chi_0$ is a constant variational parameter and
$\Delta_{i,j}$ is a nearest neighbor d-wave
pairing (with opposite signs
on the vertical and horizontal bonds) and
$\mu$ is the chemical potential.
As a matter of fact, the BCS mean-field hamiltonian can
be obtained after a mean-field decoupling of
the \tj model, where the decoupled exchange energy leads
to the $\chi_0$ and $\Delta_{i,j}$ order parameters.
In this respect, we expect that the BCS wave-function
is a good starting point to approximate the
ground state of the \tj model.
However, such a wave-function obviously does not
fulfill the constraint of no-doubly occupied site
(as in the \tj model).
This can be easily achieved, at least at the formal level,
by applying the full Gutzwiller operator~\cite{Gutzwiller}
$\mathcal{P}_\mathcal{G} =\prod_i(1-n_{i\uparrow}n_{i\downarrow})$
to the BCS wave-function $\left| {\psi_{\rm BCS} } \right\rangle$:
\begin{equation}
\label{eq:startfunc2}
    \left| \psi_{\rm RVB} \right\rangle = \PP_{\mathcal{G}}
     \left| {\psi_{\rm BCS} } \right\rangle \ .
\end{equation}
The main difficulty to deal with projected wave-functions is to treat
correctly the Gutzwiller projection $\PP_{\mathcal{G}}$.
This can be done numerically using a conceptually exact
variational Monte Carlo (VMC)
technique~\cite{yokoyama_mcv_supra,gros_mcv_supra,paramekanti_mcv}
on large clusters. It has been shown that the
magnetic energy of the variational RVB state at half-filling is
very close to the best exact estimate for the Heisenberg model.
Such a scheme also provides, at finite
doping, a semi-quantitative understanding of the phase diagram of
the cuprate superconductors and of their experimental properties.
Novel results using a VMC technique
associated to inhomogeneous wave-functions will be presented
in Section~\ref{Sec:VMC}.

Another route to deal with the Gutzwiller projection is to use a
\emph{renormalized mean-field (MF) theory}~\cite{Renormalized_MF}
in which the kinetic and superexchange energies are renormalized
by different doping-dependent factors $g_t$ and $g_J$
respectively. Further mean-field treatments of the interaction
term can then be accomplished in the particle-particle
(superconducting) channel. Crucial, now well established,
experimental observations such as the existence of a pseudo-gap
and nodal quasi-particles and the large renormalization of the
Drude weight are remarkably well explained by this early MF RVB
theory~\cite{RVB2}. An extension of this
approach~\cite{Renormalized_MF_2,Renormalized_MF_3} will be
followed in Section~\ref{Sec:MF} to investigate
inhomogeneous structures with checkerboard patterns involving a
decoupling in the particle-hole channel. As (re-) emphasized
recently by Anderson and coworkers~\cite{RVB2}, this general
procedure, via the effective MF Hamiltonian, leads to a Slater
determinant $\left| \Psi_{\rm MF}\right\rangle$ from which a
correlated wave-function $\PP_{\mathcal{G}}\left|\Psi_{\rm
MF}\right\rangle$ can be constructed and investigated by VMC.
Since the MF approach offers a reliable \emph{guide} to construct
translational symmetry-breaking projected variational
wave-functions, we will present first the MF approach in
section~\ref{Sec:MF} before the more involved VMC calculations in
Section~\ref{Sec:VMC}.

\section{Gutzwiller-projected mean-field theory}
\label{Sec:MF}
\subsection{Gutzwiller approximation and mean-field equations}

We start first with the simplest approach where the action of the
Gutzwiller projector $\PP_{\mathcal{G}}$ is approximately taken
care of using a Gutzwiller approximation scheme~\cite{Gutzwiller}.
We generalize the MF approach of
Ref.~\onlinecite{Renormalized_MF_2}, to allow for non-uniform site
and bond densities. Recently, such a procedure was followed in
Ref.~\onlinecite{Renormalized_MF_3} to determine under which
conditions a $4\times 4$ superstructure might be stable for hole
doping close to $1/8$. We extend this investigation to arbitrary
small doping and other kinds of supercells. In particular, we
shall also consider 45-degree tilted supercells such as
$\sqrt{2}\times \sqrt{2}$, $\sqrt{8}\times \sqrt{8}$ and
$\sqrt{32}\times \sqrt{32}$.

The weakly doped antiferromagnet
is described here by the renormalized
$t{-}J$ model Hamiltonian,
  \begin{equation}
   H_{t-J}^{\rm ren}= - tg_t\sum_{\langle ij\rangle\sigma}
      (c^{\dagger}_{i,\sigma}c_{j,\sigma}+h.c.)
    + Jg_J\sum_{\langle ij\rangle} \mathbf{S}_i \cdot \mathbf{S}_j
  \end{equation}
where the local constraints of no doubly occupied sites
are replaced by statistical
Gutzwiller weights $g_t=2x/(1+x)$ and $g_J=4/(1+x)^2$, where $x$ is the hole doping.
A typical value of $t/J=3$ is assumed hereafter.

Decoupling in both particle-hole and (singlet) particle-particle
channels can be considered simultaneously leading to the following
MF hamiltonian,
  \begin{eqnarray}
   H_{\rm MF}= - t \sum_{\langle ij\rangle\sigma} g_{ij}^t
      (c^{\dagger}_{i,\sigma}c_{j,\sigma}+h.c.)
+\sum_{i\sigma}\epsilon_i n_{i,\sigma}\nonumber \\
   -\frac{3}{4} J \sum_{\langle ij\rangle\sigma}g_{i,j}^J
(\chi_{ji}c^{\dagger}_{i,\sigma}c_{j,\sigma} + h.c. -|\chi_{ij}|^2)\\
   -\frac{3}{4} J \sum_{\langle ij\rangle\sigma}g_{i,j}^J
(\Delta_{ji}c^{\dagger}_{i,\sigma}c^\dagger_{j,-\sigma}
   + h.c. -|\Delta_{ij}|^2),\nonumber
  \end{eqnarray}
where the previous Gutzwiller weights have been expressed
in terms of local fugacities $z_i=2x_i/(1+x_i)$ ($x_i$ is the
local hole density $1-\langle n_i\rangle$),
$g_{i,j}^t=\sqrt{z_i z_j}$ and
$g_{i,j}^J=(2-z_i)(2-z_j)$, to allow for small non-uniform charge
modulations~\cite{Anderson_suggestion_4a}.
The Bogoliubov-de Gennes self-consistency conditions are implemented as
$\chi_{ji}=\langle c^\dagger_{j,\sigma}c_{i,\sigma}\rangle$ and
$\Delta_{ji}=\langle c_{j,-\sigma}c_{i,\sigma}\rangle
=\langle c_{i,-\sigma}c_{j,\sigma}\rangle$.

In principle, this MF treatment allows a description of modulated
phases with {\it coexisting} superconducting order, namely
supersolid phases. Previous
investigations~\cite{Renormalized_MF_3} failed to stabilize such
phases in the case of the {\it pure} 2D square lattice (i.e.
defect-free). Moreover, in this Section, we will restrict
ourselves to $\Delta_{ij}=0$. The case where both $\Delta_{ij}$
and $\chi_{ij}$ are non-zero is left for a future work, where the
effect of a defect, such as for instance a vortex, will be
studied.

In the case of finite $V_0$ , the on-site terms $\epsilon_i$
may vary spatially as $-\mu+e_i$, where $\mu$ is the chemical potential
and $e_i$ are on-site energies which are self-consistently given by,
  \begin{equation}
   e_i= \sum_{j\ne i}V_{i,j}\big<n_j\big>\, .
  \end{equation}
In that case, a constant
$\sum_{i\ne j}V_{i,j}(\langle n_i\rangle \langle n_j\rangle+n^2)$ has to be added
to the MF energy.
Note that we assume here a {\it fixed} chemical potential $\mu$.
In a recent work \cite{chemical_function}, additional degrees of freedom
where assumed (for $V_0=0$) implementing an unconstrained minimization
with respect to the on-site fugacities. However, we believe
that the energy gain is too small to be really conclusive
(certainly below the accuracy one can expect from such a simple MF approach).
We argue that we can safely neglect the spatial variation of $\mu$ in first
approximation, and this will be confirmed by the more accurate
VMC calculations in Section~\ref{Sec:VMC}.
Incidently, Ref.~\onlinecite{chemical_function} emphasizes
a deep connection between the stability of
checkerboard structures~\cite{Renormalized_MF_3} and the instability
of the SFP due to nesting properties of some parts of its
Fermi surface \cite{susceptibility}.

\subsection{Mean-field phase diagrams}

In principle, the mean-field equations could be solved in real
space on large clusters allowing for arbitrary modulations of the
self-consistent parameters. In practice, such a procedure is not
feasible since the number of degrees of freedom involved is too
large. We therefore follow a different strategy. First, we assume
fixed (square shaped) supercells and a given symmetry within the
super-cell (typically invariance under 90-degrees rotations) to
reduce substantially the number of parameters to optimize.
Incidently, the assumed periodicity allows us to conveniently
rewrite the meanfield equations in Fourier space using a reduced
Brillouin zone with a very small mesh. In this way, we can
converge to either an absolute or a local minimum. Therefore, in a
second step, the MF energies of the various solutions are compared
in order to draw an overall phase diagram.

\begin{figure}
  \begin{center}
    \includegraphics[width=\midwidth]{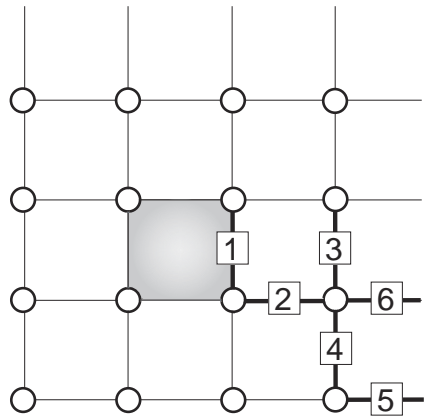}
    \caption{$4\times 4$ unit-cell used in both the MF approach and
the variational wave-function. Note the existence of 6 independent
bonds (bold bonds), that for convenience are labelled from $1-6$, and of
3 {\it a priori} non-equivalent sites. The center of the dashed
    plaquette is the center of the (assumed) $C_{4V}$ symmetry. Other
    sizes of the same type of structure are considered in the MF case,
    respectively : $2 \times 2$, $\sqrt{8} \times \sqrt{8}$, and
    $\sqrt{32} \times \sqrt{32}$ unit cells.}
      \label{cell}
  \end{center}
\end{figure}

In previous MF calculations~\cite{Renormalized_MF_3},
stability of an inhomogeneous solution with the
$4\times 4$ unit-cell shown in Fig.~\ref{cell} was found
around $x=1/8$. Here, we investigate its stability for arbitrary doping
and extend the calculation to another possible competing
solution with a twice-larger (square) unit-cell containing 32 sites.
The general solutions with different
phases and/or amplitudes on the independent links will be refered
to as bond-order (BO) phases~\cite{note_charge}.
Motivated by experiments~\cite{STM-BSCO,STM2},
a $C_{4V}$ symmetry of the inhomogeneous patterns around a central
plaquette will again be assumed for both cases.
Note that such a feature greatly reduces the number
of variational parameters and hence speeds up the convergence
of the MF equations. Starting from a central plaquette,
like in Fig.~\ref{cell}, a larger
$\sqrt{32}\times\sqrt{32}$ unit-cell (not shown)
can easily be constructed with
10 non-equivalent bonds (with both independent
real and imaginary parts) and {\it a priori} 6 non-equivalent sites.
Note that this new unit-cell is now tilted by 45 degrees.

\begin{figure}
  \begin{center}
    \includegraphics[width=\midwidth]{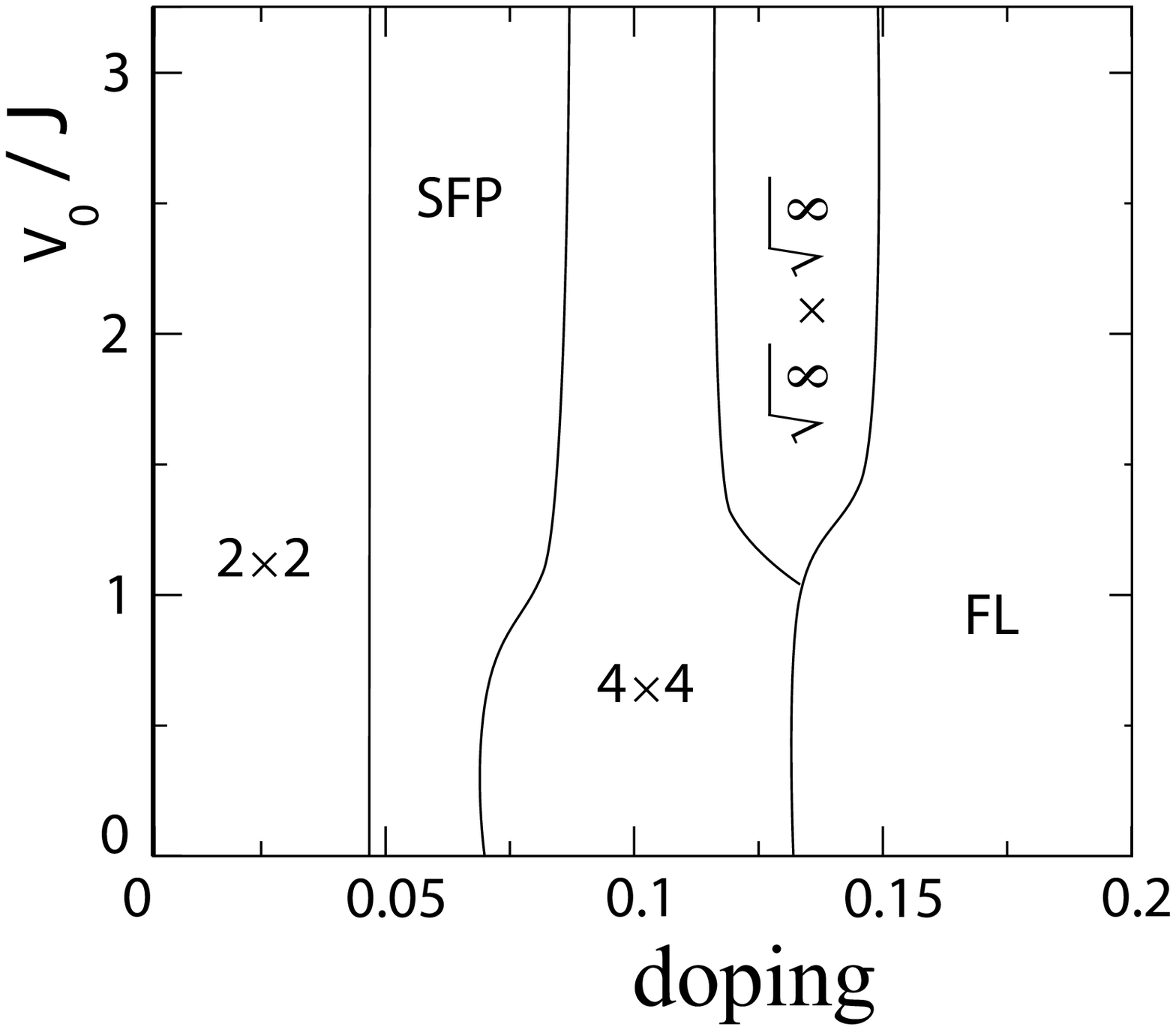}
    \includegraphics[width=\midwidth]{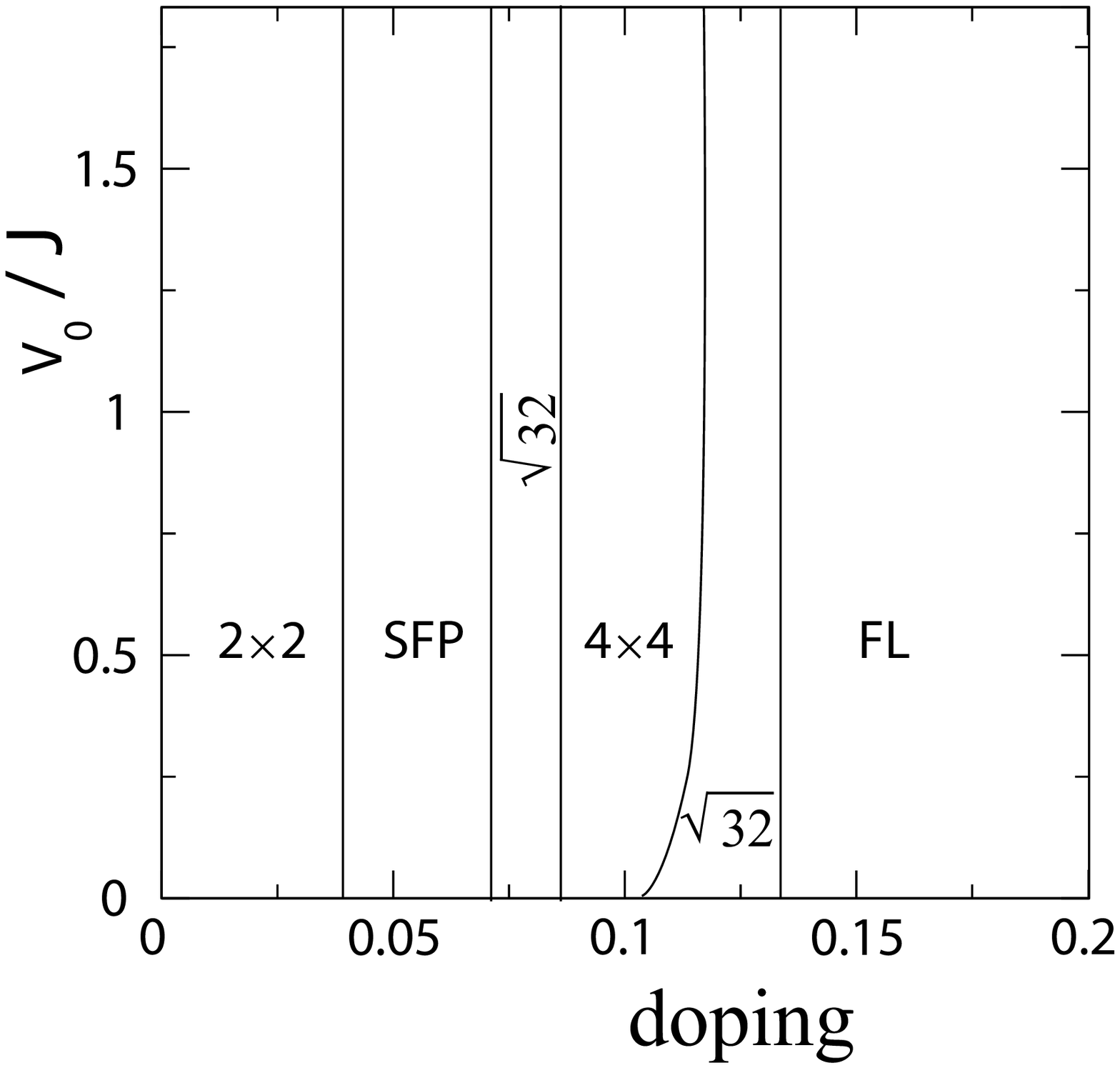}

    \caption{Mean-field phase diagrams obtained by solving self-consistently
the mean-field equations on a $128 \times 128$ lattice
(for $\ell_0=4$) vs hole doping $x$ and repulsion $V_0$ (in units of $J$).
Top: results obtained assuming
a $4\times 4$ unit cell; Bottom: same with a $\sqrt{32}\times
\sqrt{32}$ tilted unit cell.
In both cases, a $C_{4v}$ symmetry is assumed (see text).}

      \label{meanfield_phase}
  \end{center}
\end{figure}

At this point, it is important to realize that patterns with a
smaller number of non-equivalent bonds or sites are in fact
subsets of the more general modulated structures described above.
For example, the SFP is obviously a special case of such patterns,
where all the $\chi_{i,j}$ are equal in magnitude with a phase
oriented to form staggered currents, and where all the sites are
equivalent. This example clearly indicates that the actual
structure obtained after full convergence of the MF equations
could have {\it higher} symmetry than the one postulated in the
initial configuration which assumes a random choice for all
independent parameters. In particular, the equilibrium unit-cell
could be smaller than the original one and contain a fraction
($1/2$ or $1/4$) of it. This fact is illustrated in
Fig.~\ref{meanfield_phase} showing two phase diagrams produced by
using different initial conditions, namely $4 \times 4$ (top) and
$\sqrt{32} \times \sqrt{32}$ (bottom) unit-cells. Both diagrams
show consistently the emergence of the SFP at dopings around $6\%
$ and a \emph{plaquette} phase ($2 \times 2$ unit-cell with two
types of bonds) at very small doping~\cite{note_AF,Vojta}. In
addition, a phase with a $\sqrt{8}\times\sqrt{8}$ super-cell is
obtained for a specific range of doping and $V_0$ (see
Fig.~\ref{meanfield_phase} on the top). Interestingly enough, all
these BO phases can be seen as {\it bond-modulated SFP} with 2, 4,
8 and up to 16 non-equivalent (staggered) plaquettes of slightly
different amplitudes. This would be consistent with the SFP
instability scenario~\cite{susceptibility} which suggests that the
wavevector of the modulation should vary continuously with the
doping. Although this picture might hold when $V_0=0$, our
results show that the system prefers some peculiar spatial
periodicities (like the ones investigated here) that definitely
take place at moderate $V_0$.

Let us now compare the two phase diagrams.
We find that, except in some doping regions,
the various solutions obtained with the $4\times 4$ unit-cell are
recovered starting from a twice larger unit-cell.
Note that, due to the larger number of parameters,
the minimization procedure starting from a larger unit-cell
explores a larger phase space and it is expected to be more likely
to converge
to the absolute minimum. This is particularly clear
(although not always realized) at large doping $x=0.14$, where we
expect an homogeneous Fermi Liquid (FL) phase (all bonds are real and equal),
as indeed seen in Fig.~\ref{meanfield_phase} on the bottom.
On the contrary, Fig.~\ref{meanfield_phase} on the top reveals,
for $V_0/J \in[1.5,3]$,
a modulated $\sqrt{8}\times\sqrt{8}$ structure, which is an artefact
due to the presence of a local minimum (see next).

\begin{figure}
  \begin{center}
    \includegraphics[width=1.\figwidth]{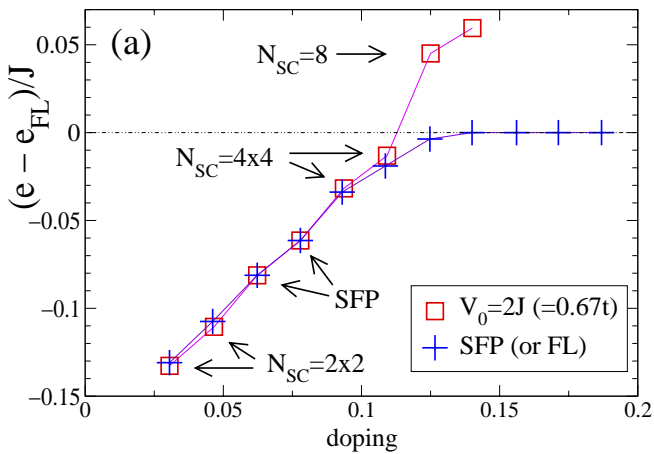}
\end{center}
\begin{center}
    \includegraphics[width=1.\figwidth]{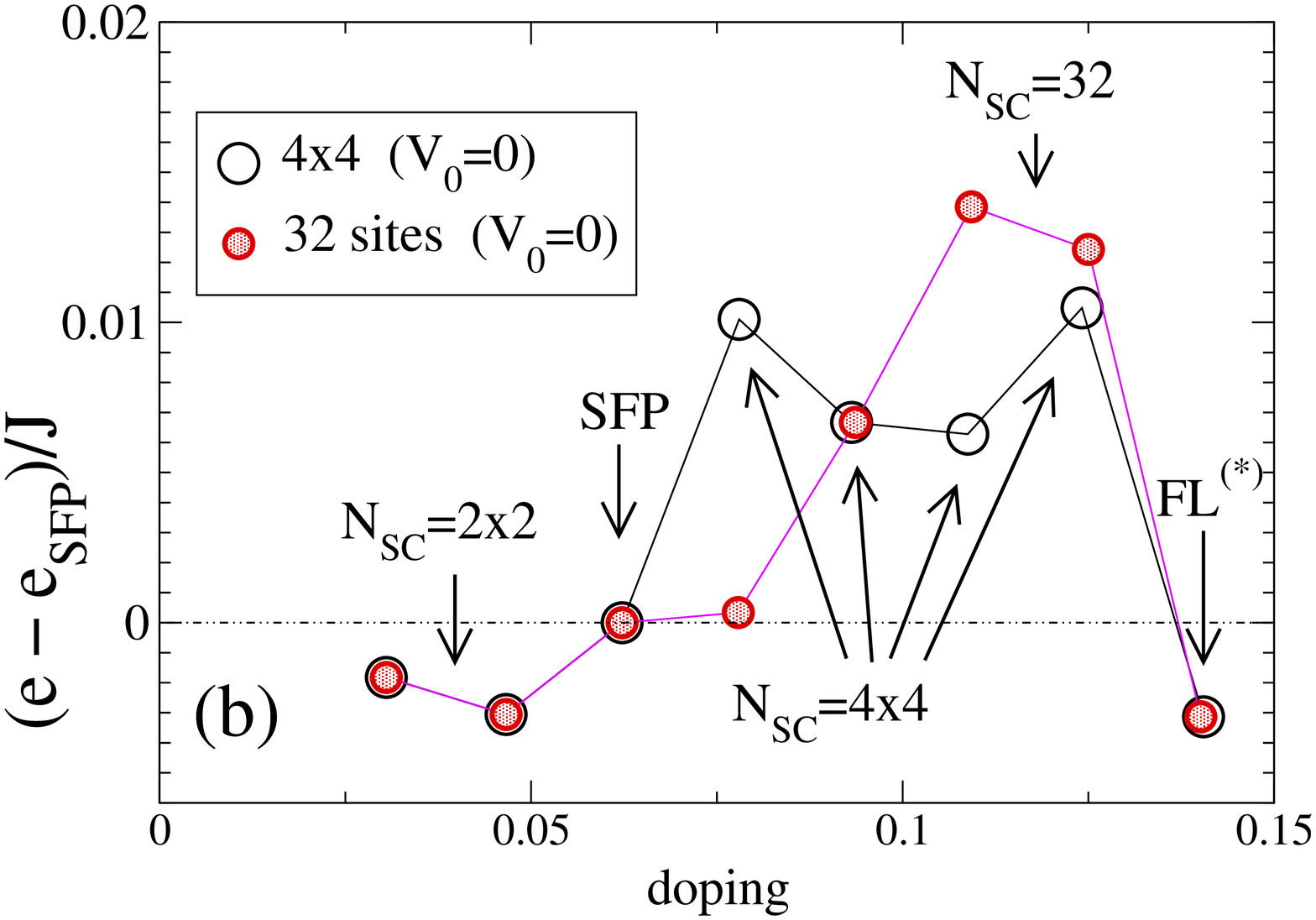}
    \caption{(Color online) (a) Energy per site (in units of $J$ and for
$t=3J$) obtained by solving the mean-field equations using the
{\it initial} $4\times 4$ unit-cell (see text) for a moderate
value of $V_0$. The SFP energy is also shown for comparison. The
FL energy has been substracted from all data for clarity. (b)
Comparison of the energies (for $V_0=0$) using different initial
conditions (see text), a $4\times 4$ or a
$\sqrt{32}\times\sqrt{32}$ unit-cell; due to very small energy
differences, the SFP energy is used as a reference for an easier
comparison. The different phases specified by arrows and
characterized by the number of sites $N_{SC}$ of their actual
supercells refer to the ones in
Fig.~\protect\ref{meanfield_phase}. For doping $x=0.14$, the
minimization leads to a solution with small imaginary parts (of
order $10^{-4}$) very similar to a FL phase, which we call FL$^*$.
}
      \label{meanfield_energy}
  \end{center}
\end{figure}

Since the MF procedure could accidentally give rise to local
minima, it is of interest to compare the MF energies obtained by
starting with  \emph{random} values of all independent parameters
within the two previously discussed unit-cells. For convenience,
we have substracted from all data either the FL (in
Fig.~\ref{meanfield_energy}(a)) or the SFP (in
Fig.~\ref{meanfield_energy}(b)) \emph{reference} energy. From
Figs.~\ref{meanfield_energy}(a,b) we see that we can converge
towards a local energy minimum, often modulated in space, which is
not the absolute minimum. Indeed, over a large doping range, the
lowest energy of the all solutions we have found is obtained for
homogeneous densities and bond magnitudes. Nevertheless, we see
that the $4\times 4$ modulated phase is (i) locally stable and
(ii) is very close in energy to the homogeneous (SFP) phase which,
often, has a slightly lower energy. Note that, around $x\simeq
1/8$, the states with $\sqrt{8}\times\sqrt{8}$ and
$\sqrt{32}\times\sqrt{32}$ supercells are clearly metastable
solutions (and using a larger initial unit-cell is not favorable
in the latter case). In contrast,  in this range of doping, the
$4\times 4$ checkerboard state is very competitive w.r.t. the SFP.
Therefore, it makes it a strong candidate to be realized either in
the true ground state of the model, or present as very low excited
state~\cite{Vojta2}. In fact, considering such small energy
differences, it is clear that an accurate comparison is beyond the
accuracy of the MF approach. We therefore move to the
\emph{approximation-free} way of implementing the Gutzwiller
projection with the VMC technique, that allows a detailed
comparison between these variational homogeneous and inhomogeneous
states.

\section{Variational Monte Carlo simulations of $4\times 4$ superstructures}
\label{Sec:VMC}

Motivated by the previous mean-field results we have carried out
extensive Variational Monte Carlo simulations. In this approach,
the action of the Gutzwiller projection operator is taken care of
exactly, although one has to deal with finite clusters. In order
to get rid of discontinuities in the d-wave RVB wave-function, we
consider (anti-)periodic boundary conditions along $e_y$ ($e_x$).
As a matter of fact, it is also found that the energy is lower for
twisted boundary conditions, hence confirming the relevance of
this choice of boundaries. We have considered a $16 \times 16$
square cluster of $N=256$ sites. We also focus on the $1/8$ doping
case which corresponds here to $N_e=224$ electrons on the 256 site
cluster. Following the previous MF approach, we consider the same
generic mean-field hamiltonian, \begin{equation}
  H_{\rm MF}= \sum\limits_{\langle i,j \rangle,\sigma}
  \left( -\ {\tilde t}_{i,j} c_{i\sigma }^{\dagger} c_{j\sigma} + h.c.\right)
  +\sum\limits_{i\sigma} \epsilon_i c_{i\sigma }^{\dagger} c_{i\sigma}\, ,
  \label{MF_hamilt}
\end{equation}
where the complex bond amplitudes ${\tilde t}_{i,j}$ can be written as
$\left| {\tilde t}_{i,j}\right| e^{i{\theta}_{i,j}}$, and
${\theta}_{i,j}$ is a phase oriented on the bond $i \to j$. The on-site
terms $\epsilon_i$ allow to control the magnitude of the charge
density wave. However, the energy was found to be minimized for
all the $\epsilon_i$ equal to the same value in the range
$V_0=[0,5]$ and for the two parameters $\ell_0=2,4$. In fact, we find
that strong charge ordered wave-functions are not stabilized in
this model~\cite{note_charge}.

In this Section, we shall restrict ourselves to the $4 \times 4$
unit-cell where all independent variational parameters are to be
determined from an energy minimization. This is motivated both by
experiments~\cite{STM-BSCO,STM2} and by the previous MF results
showing the particular stability of such a structure (see also
Ref.~\onlinecite{Renormalized_MF_3}). As mentioned in the previous
Section, we also impose that the phases and amplitudes respect the
$C_{4V}$ symmetry within the unit-cell (with respect to the center
of the middle plaquette, see Fig.~\ref{cell}), reducing the
numbers of independent links to $6$. To avoid spurious
degeneracies of the MF wave-functions related to multiple choices
of the filling of the discrete k-vectors in the Brillouin Zone (at
the Fermi surface), we add very small random phases and amplitudes
($10^{-6}$) on all the links in the $4 \times 4 $ unit cell.

Let us note that commensurate flux phase (CFP) are also candidate
for this special $1/8$ doping. In a previous study, a subtle
choice of the phases ${\theta}_{i,j}$ (corresponding to a gauge
choice in the corresponding Hofstadter problem~\cite{Hofstadter})
was proposed \cite{Flux_phases_1}, which allows to write the
${\phi}=p/16~(p<16)$ flux per plaquette wave-function within the
same proposed unit-cell~\cite{Flux_phases_1} and is also expected
to lead to a better kinetic energy than the Landau gauge (in the
Landau gauge the unit-cell would be a line of $16$ sites).
However, we have found that the CFP wave-functions turned out
\emph{not} to be competitive for our set of parameters $V_0$, due
to their quite poor kinetic energy, although they have very good
Coulomb and exchange energies. We argue that such CFP
wave-functions would become relevant in the large Coulomb and/or
$J$ regimes (see table~\ref{energies}).

In order to further improve the energy, we also add a
nearest-neighbor spin-independent Jastrow \cite{Jastrow_factor}
term to the wave-function,
\begin{equation}\label{eq:jastrow}
  \PP_{\mathcal{J}}=\exp{\left( \alpha\sum\limits_{ \langle i,j \rangle }{n_i n_j }\right)},
\end{equation}
where $\alpha$ is an additional variational parameter. Finally,
since the $t-J$ model allows at most one fermion per site, we
discard all configurations with doubly occupied sites by applying
the complete Gutzwiller projector $\PP_\mathcal{G}$. The
wave-function we use as an input to our variational study is therefore
given by,
\begin{equation}
\label{eq:startfunc}
    \left| \psi_{\rm var} \right\rangle = \PP_{\mathcal{G}} \PP_{\mathcal{J}}
     \left| {\psi _{\rm MF} } \right\rangle
\end{equation}
In the following, we shall introduce simple notations for denoting
the various variational wave-functions, $BO$ for the bond-order
wave function, $SFP$ for the staggered flux phase, $RVB$ for the
d-wave RVB superconducting phase, $FS$ for the simple projected
Fermi-Sea, and we will use the notation $MF/\mathcal{J}$
($MF=BO,SFP,RVB,FS$) when the Jastrow factor is applied on the
mean-field wave-function. Finally, it is also convenient to
compare the energy of the different wave-functions with respect to
the energy of the simple projected Fermi-Sea (i.e. the correlated
wave-function corresponding to the previous FL mean-field phase),
therefore we define a \emph{condensation energy} as
$e_c=e_{var}-e_{FS}$.

\begin{table}
\caption{Set of energies per lattice site for $V_0=1$ and
$\ell_0=4$ for different wave-functions. The best commensurate
flux phase in the Landau gauge with flux per plaquette $p/16$ was
found for $p=7$. We also show the energy of the CFP with flux
$7/16$ written with another choice of gauge. We show the total
energy per site ($E_{\rm{tot}}$), the kinetic energy per site
($E_{\rm{T}}$), the exchange energy per site ($E_{\rm{J}}$) and
the Coulomb energy per site ($E_{\rm{V}}$).} \label{energies}
\begin{tabular}{|c|c|c|c|c| }
  \hline
  \hline
  wave-function & $E_{\rm{tot}}$ & $E_{\rm{T}}$ & $E_{\rm{J}}$ & $E_{\rm{V}}$ \\
  \hline
  $FS$ & -0.4486(1) & -0.3193(1) &-0.1149(1) & -0.0144(1) \\
  $CFP~7/16^1$ & -0.3500(1) & -0.1856(1) & -0.1429(1) & -0.0216(1)\\
  $CFP~7/16^2$ & -0.4007(1) & -0.2369(1) & -0.1430(1) & -0.0208(1)\\
  $SFP$ & -0.4581(1) & -0.3106(1) & -0.1320(1) & -0.0155(1) \\
  $BO$ & -0.4490(1) & -0.3047(1) & -0.1302(1) & -0.0141(1) \\
  $RVB$ & -0.4564(1) & -0.3080(1) & -0.1439(1) & -0.0043(1) \\
  $SFP/\mathcal{J}$ & -0.4601(1) & -0.3116(1) & -0.1315(1) & -0.0169(1) \\
  $BO/\mathcal{J}$ &  -0.4608(1) & -0.3096(1) & -0.1334(1) & -0.0177(1) \\
  $RVB/\mathcal{J}$ & -0.4644(1) & -0.3107(1) & -0.1440(1) & -0.0086(1)\\
  \hline
  \hline
\end{tabular}
\begin{flushleft}
$^1$ Landau gauge \linebreak
$^2$ Gauge of Ref.\protect\onlinecite{Flux_phases_1}
\end{flushleft}
\end{table}

\begin{table}
\caption{Order parameters for the different wave-functions for
$V_0=1.5$ and $\ell_0=4$. We depict the following order
parameters: $t_{i,j} \times e^{i \phi_{i,j}}$, where $t_{i,j}$
($\phi_{i,j}$) is the amplitude (phase) of $\langle c_i^+ c_j
\rangle$, and the exchange energy $\langle S_i.S_j\rangle$, for
the 6 independent bonds labelled for convenience according to Fig.
\ref{cell}.  The sign of $\phi_{i,j}$ is according to the
staggered flux pattern (see arrows in Fig. \ref{snapshots}). We
note that the $RVB/\mathcal{J}$ is uniform by construction. The
variational superconducting order parameter is $\Delta_{RVB}=0.3$
for the $RVB/\mathcal{J}$ wave-function and $\Delta_{RVB}=0$ for
the $SFP/\mathcal{J}$ and $BO/\mathcal{J}$ wave-functions.}
\label{data}
\begin{center}
\begin{tabular}{|c|c|c|c|c|c|c| }
  \hline
  \hline
  \phantom{aaa} & bond 1 & bond 2 & bond 3 & bond 4 & bond 5 & bond 6 \\
  \hline
  $t_{i,j}$ &&&&&&\\
  $RVB/\mathcal{J}$ & 0.077(1)  & 0.077(1)  & 0.077(1)   & 0.077(1)  & 0.077(1)   & 0.077(1) \\
  $SFP/\mathcal{J}$ & 0.085(1)  & 0.085(1)  & 0.085(1)   & 0.085(1)  & 0.085(1)   & 0.085(1) \\
  $BO/\mathcal{J}$  & 0.082(1)  & 0.083(1)  & 0.093(1)   & 0.088(1)  & 0.086(1)   & 0.084(1) \\
  &&&&&&\\
  $\left| \phi_{i,j} \right|$ &&&&&& \\
  $RVB/\mathcal{J}$ & 0 & 0 & 0 & 0 & 0 & 0 \\
  $SFP/\mathcal{J}$ & 0.438(1) & 0.438(1) & 0.438(1)  & 0.438(1)  & 0.438(1)  & 0.438(1)  \\
  $BO/\mathcal{J}$  & 0.527(1) & 0.502(1) & 0.473(1)  & 0.390(1)  & 0.338(1)  & 0.384(1)  \\
  &&&&&&\\
  $-\langle S_i.S_j \rangle$ &&&&&& \\
  $RVB/\mathcal{J}$ & 0.215(1) & 0.215(1)  & 0.215(1) &0.215(1)  &0.215(1)  & 0.215(1) \\
  $SFP/\mathcal{J}$ & 0.197(1) & 0.197(1)  &0.197(1)  &0.197(1)  &0.197(1)  &0.197(1)  \\
  $BO/\mathcal{J}$  & 0.215(1) & 0.207(1) &0.215(1) &0.187(1) &0.186(1) &0.170(1) \\
  \hline
  \hline
\end{tabular}
\end{center}
\end{table}

\begin{figure}
  \begin{center}
    \includegraphics[width=\figwidth]{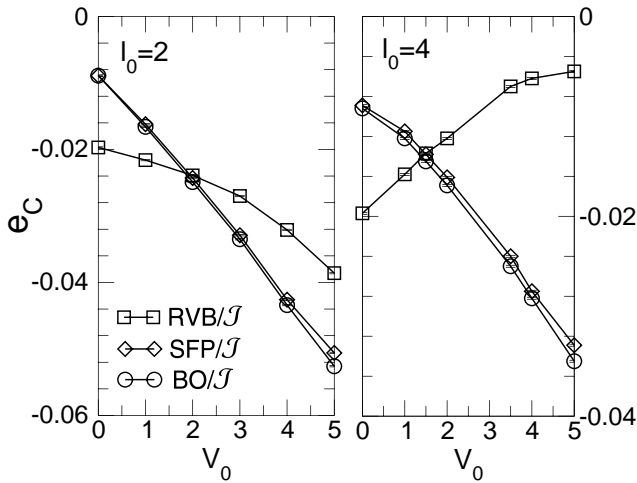}
    \caption{Energy per lattice site of the
     $RVB/\mathcal{J}$, $SFP/\mathcal{J}$ and $BO/\mathcal{J}$ wave-functions
    minus the energy of the projected Fermi-Sea wave-function.
    }
      \label{ec}
  \end{center}
\end{figure}

\begin{figure}
  \begin{center}
    \includegraphics[width=\figwidth]{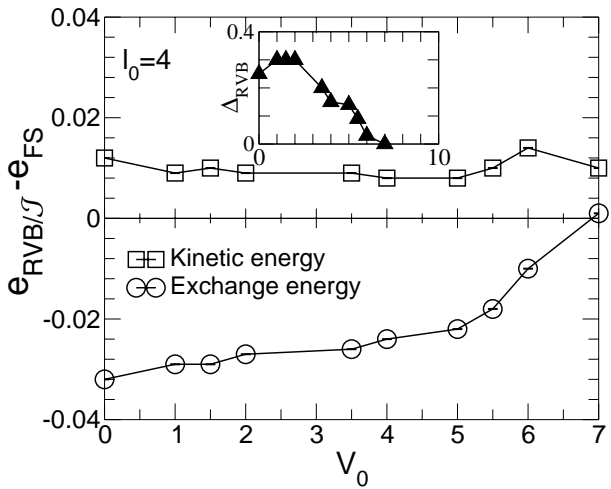}
    \caption{Kinetic and exchange
    energy per site of the $RVB/\mathcal{J}$ wave-function minus the respective
    exchange and kinetic energy of the simple projected
    Fermi-Sea. Inset: value of the variational d-wave gap.
    }
      \label{RVBcipo}
  \end{center}
\end{figure}

In Fig.~\ref{ec} we present the energies of the three
wave-functions $BO/\mathcal{J}$, $SFP/\mathcal{J}$ and $RVB/\mathcal{J}$ for Coulomb potential
$V_0\in [0,5]$. We find that for both $\ell_0=2$ and $\ell_0=4$ the
$RVB$ phase is not the best variational wave-function when the
Coulomb repulsion is strong. The bond-order wave-function has a
lower energy for $V_0>2$ and $\ell_0=2$ ($V_0>1.5$ and $\ell_0=4)$. Note
that the (short range) Coulomb repulsion in the cuprates is
expected to be comparable to the Hubbard $U$, and therefore
$V_0\approx 5$ or $10$ seems realistic. Independently of the
relative stability of both wave-functions, the superconducting
d-wave wave-function itself is strongly destabilized by the
Coulomb repulsion as indicated by the decrease of the variational
gap parameter for increasing $V_0$ and the suppression of
superconductivity at $V_0\simeq 7$ (see Fig.~\ref{RVBcipo}).

\begin{figure}
  \begin{center}
    \includegraphics[width=\figwidth]{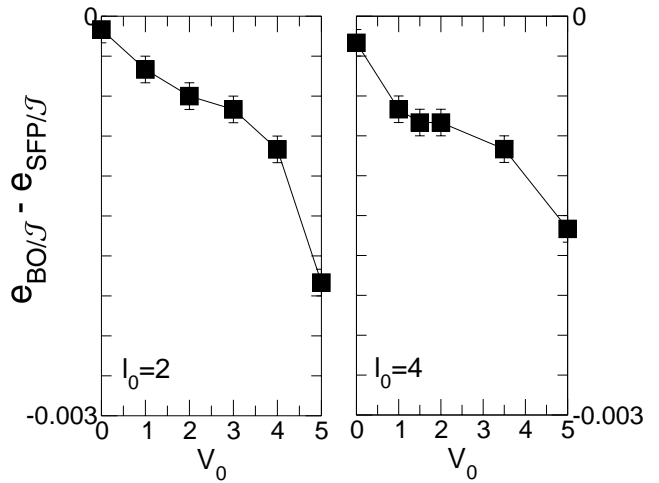}
    \caption{Total energy per site of the $BO/\mathcal{J}$ minus the
    energy of the $SFP/\mathcal{J}$ wave-functions.
    }
      \label{BO_SFP_etot}
  \end{center}
\end{figure}
\begin{figure}
  \begin{center}
    \includegraphics[width=\figwidth]{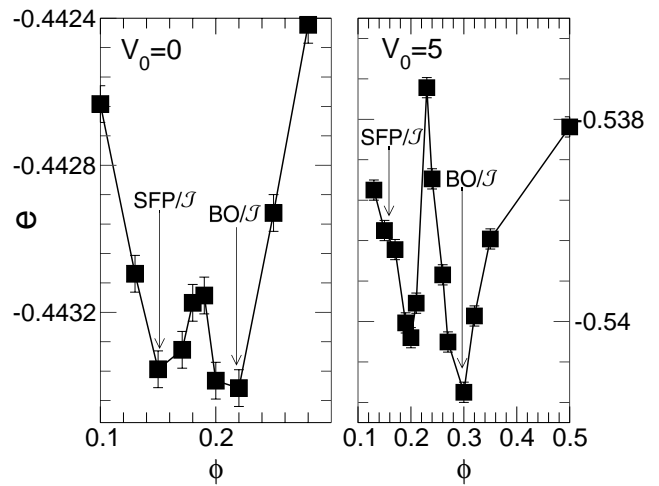}
    \caption{Total energy per site of the $BO/\mathcal{J}$ variational wave-function
    with variational parameters $Im \left({\tilde t}_{i,j}\right)=\pm \phi$
    on the bonds $1,2,3$,
    and $Im \left({\tilde t}_{i,j}\right)=\pm 0.149$ on the bonds $4,5,6$.
    The sign of $Im \left({\tilde t}_{i,j}\right)$ is oriented according to
the staggered flux pattern. We have chosen for all the links $Re
\left(\ {\tilde t}_{i,j}\right)=0.988$. Results for $V_0=0$ and $V_0=5$
with $\ell_0=4$ are shown.
    }
      \label{barrier}
  \end{center}
\end{figure}
\begin{figure}
  \begin{center}
    \includegraphics[width=\figwidth]{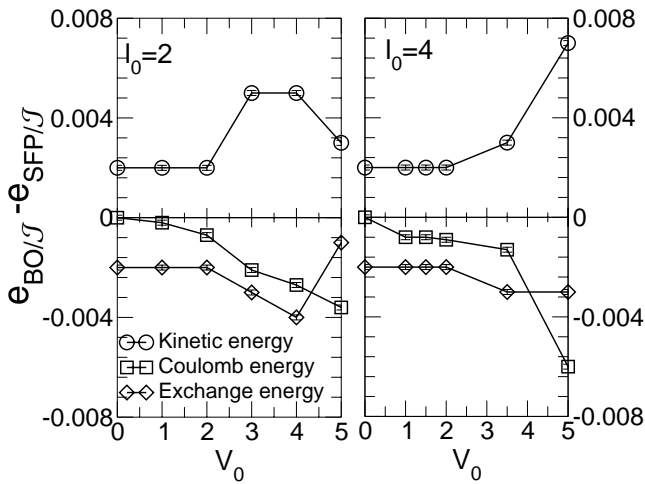}
    \caption{Kinetic, exchange and Coulomb
    energy per site of the $BO/\mathcal{J}$ wave-function minus the respective
    associated energy of the $SFP/\mathcal{J}$ wave-function.
    }
      \label{sfp_bo}
  \end{center}
\end{figure}

\begin{figure*}
\begin{tabular}{|c|c|c|c|}
 \hline
 \hline
    \includegraphics[width=\smallwidth]{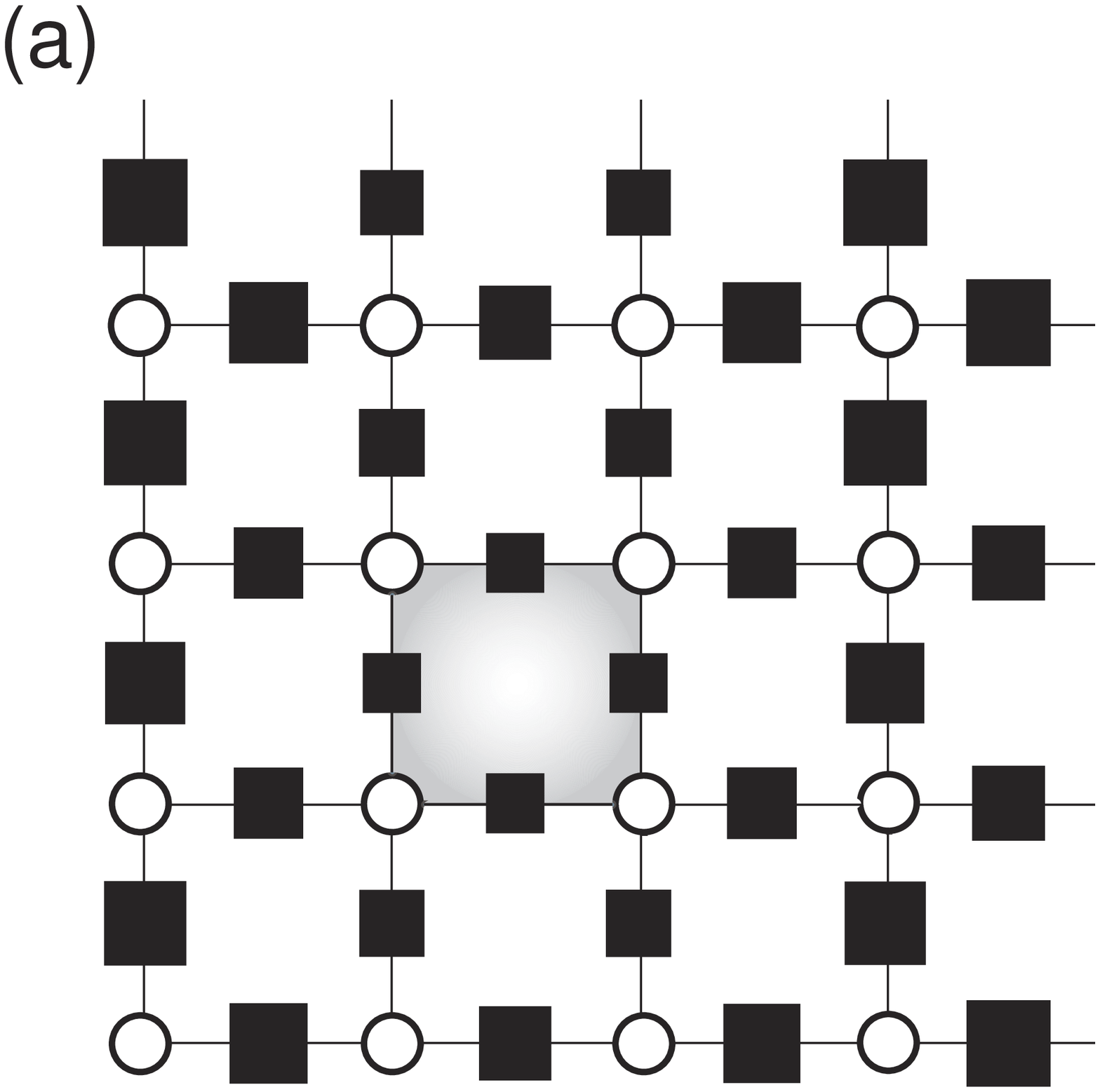}
      &
    \includegraphics[width=\smallwidth]{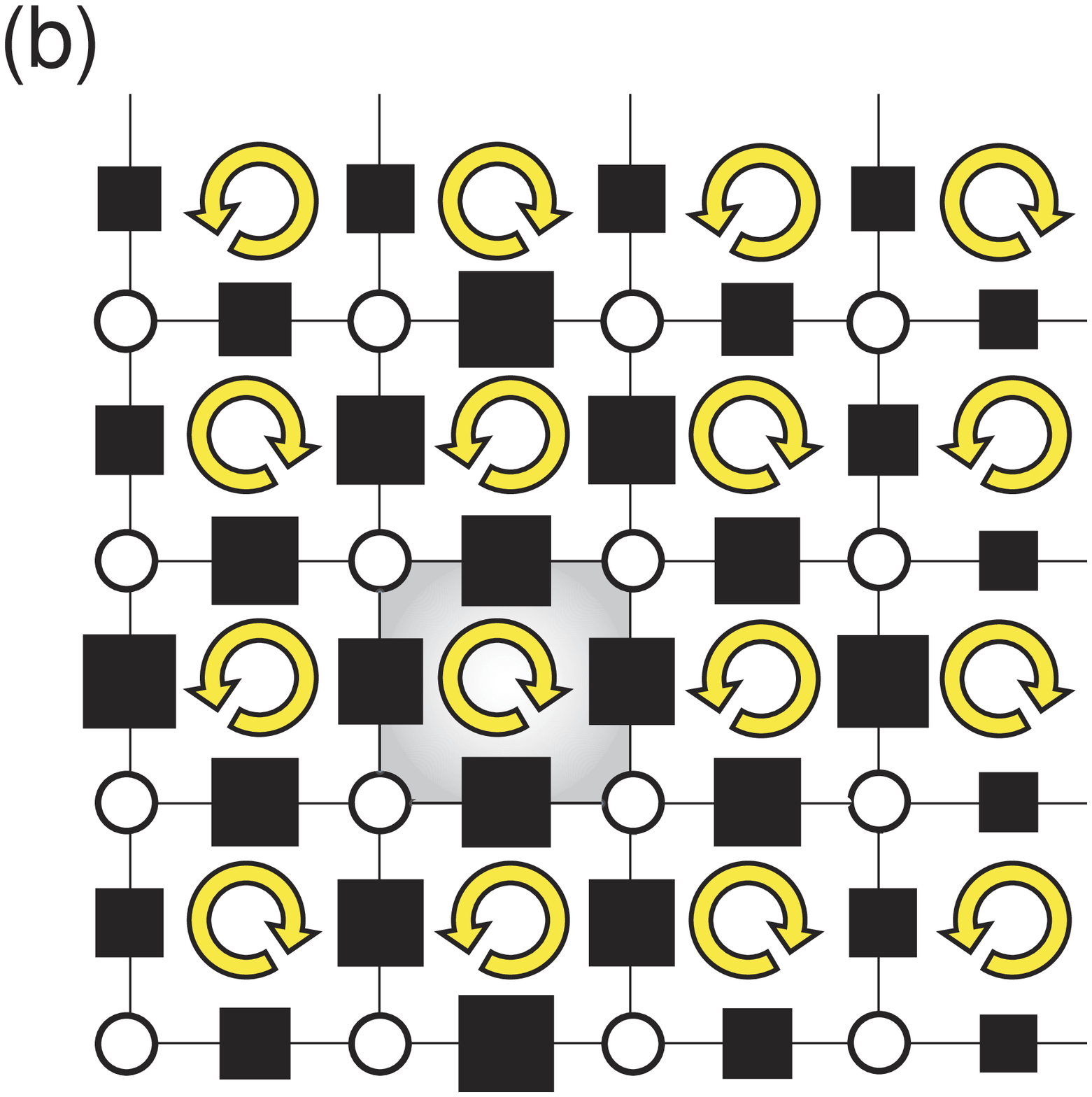}
      &
    \includegraphics[width=\smallwidth]{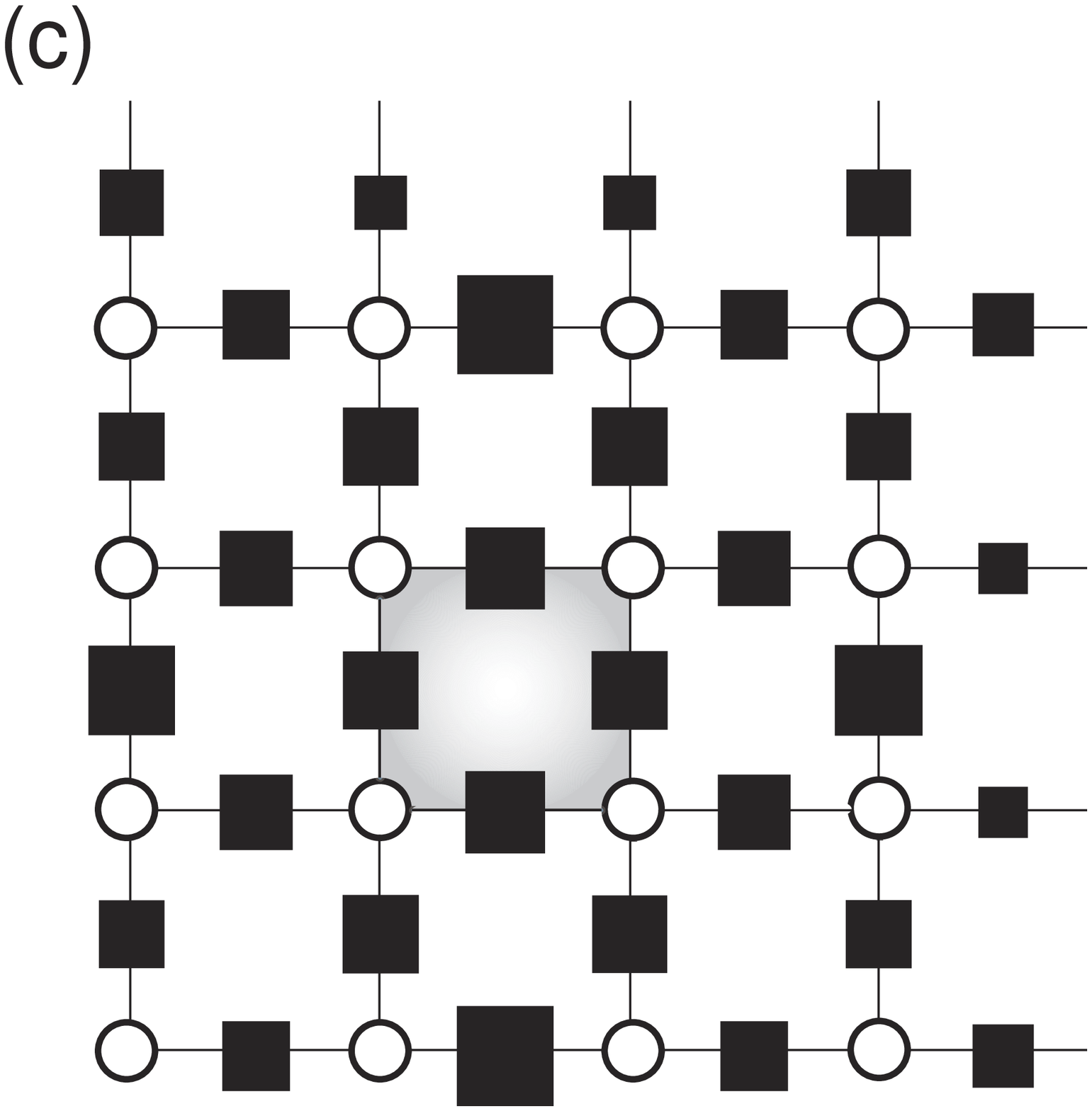}
      &
    \includegraphics[width=\smallwidth]{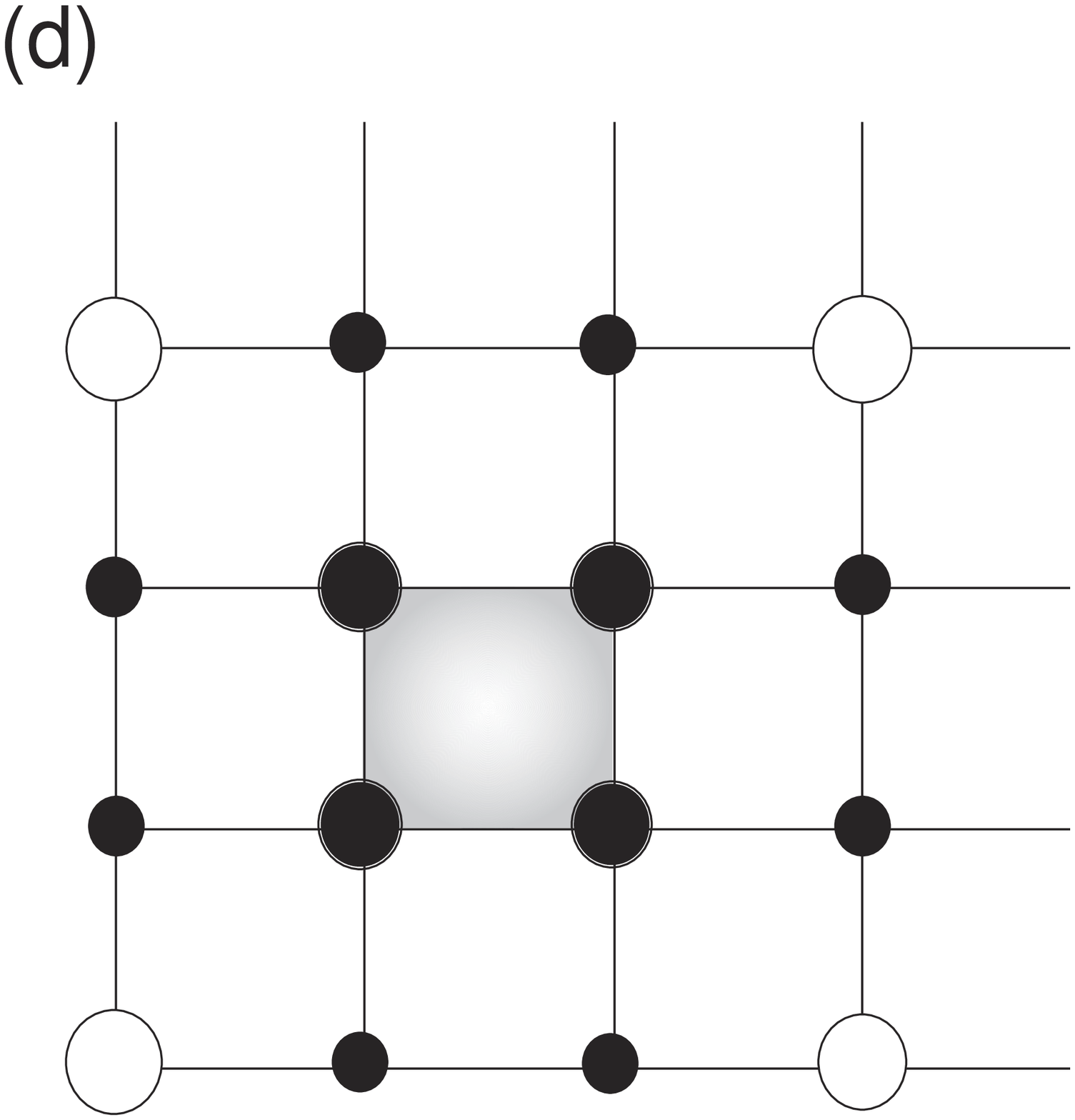}
\\
\hline \hline
\end{tabular}
 \caption{(Color online) Local expectation values (a,b,c) of the kinetic
 and exchange energies of the projected $BO/\mathcal{J}$ wave-functions
 on each of the bonds within
 the unit-cell. Width of filled square symbols is proportional to the
 (a) real  and (b) imaginary part of $\langle c_{i}^{+}c_j \rangle$,
 and (c) to the local exchange energy $\langle\mathbf{S}_i\cdot \mathbf{S}_j
\rangle$.
 The sign of the imaginary part of the hopping bonds is
 according to the staggered flux pattern (arrows).
 The wave-function has small charge density variations (d),
 therefore we subtract the mean value $n$ to the local density:
 size of circles are proportional to $\langle n_i-n\rangle$, and circles
 are open (filled) for negative (positive) sign. The biggest
 circle corresponds to an on-site charge deviation of $2\%$.
 All the above results are for $\l_0=4$ and $V_0=5$.
 }
 \label{snapshots}
\end{figure*}

Nevertheless, we observe that the difference in energy between the
bond-order wave-function and the staggered flux phase remains
very small. We show in table \ref{data} the order
parameters measured after the projection for the
$RVB/\mathcal{J}$, $SFP/\mathcal{J}$ and $BO/\mathcal{J}$
wave-functions. As expected the $RVB/\mathcal{J}$ and the
$SFP/\mathcal{J}$ wave-functions are homogenous within the
unit-cell. In contrast, the  $BO/\mathcal{J}$ wave-function shows
significant modulations (expected to be measurable experimentally)
of the various bond variables w.r.t their values
in the homogeneous SFP.
In Fig.~\ref{BO_SFP_etot} we show the small energy difference (see scale)
between the two wave-functions. Interestingly, the difference is
increasing with the strength of the potential. We notice that
the two wave-functions correspond to two nearby local minima of
the energy functional at zero Coulomb potential (see
Fig.~\ref{barrier}), which are very close in energy (the
$BO/\mathcal{J}$ wave-function is slightly lower in energy than
the $SFP/\mathcal{J}$) and are separated by a quite small energy barrier.
Note that in Fig.~\ref{barrier} we consider the variational bond
order parameters and not the projected quantities.

When the repulsion is switched on, the height of the
energy barrier increases and the $SFP/\mathcal{J}$ wave-function
does not correspond anymore to the second local minima. Indeed,
when $V_0>0$ the second local energy minima \emph{shifts}
continuously from the point corresponding to the simple
$SFP/\mathcal{J}$ wave-function. The metastable wave-function
lying at this second local minima is a weak bond-order
(SFP-like) wave-function that preserves better the
large kinetic energy while still
being able to optimize better the Coulomb energy than the homogeneous
SFP. Moreover, to understand better the stabilization of
the BO-modulated staggered flux wave-functions w.r.t the
homogeneous one, we have plotted in  Fig.~\ref{sfp_bo}
the difference in the respective kinetic energy, the exchange
energy and the Coulomb energy of the $SFP/\mathcal{J}$ and
$BO/\mathcal{J}$ wave-functions. We
conclude that the two wave-functions, although
qualitatively similar (they both exhibit an underlying staggered flux
pattern), bear quantitative differences:
the staggered flux phase (slightly) better optimizes the kinetic energy
whereas the bond-order wave-function (slightly)
better optimizes the Coulomb and exchange energies so that a small overall
energy gain is in favor of the modulated phase.
Therefore, we unambiguously  conclude that, generically, bond-order
modulations should spontaneously appear on top of the
staggered flux pattern for moderate doping.

Finally, we emphasize that the bond-order wave-function is not
stabilized by the Coulomb repulsion alone (like for a usual
electronic Wigner cristal) exhibiting coexisting bond order and
(small) charge density wave. Moreover, the variational parameters
$\epsilon_i$ in Eq.~(\ref{MF_hamilt}) are found after minimizing
the projected energy to be set to equal values on every site of
the unit-cell. Let us also emphasize that the bond-order
wave-function is not superconducting as proposed in some
scenarios~\cite{Anderson_suggestion_4a}. In the actual variational
framework, we do not consider bond-order wave-function embedded in
a sea of d-wave spin singlet pairs.

In fact, we do not expect a bulk d-wave RVB state to be stable at
large Coulomb repulsion (because of its very poor Coulomb energy)
nor a bulk {\it static} checkerboard SFP at too small Coulomb
energy. However, for moderate Coulomb repulsion for which the
d-wave RVB remains globally stable, sizeable regions of
checkerboard SFP could be easily nucleated e.g. by defects. This
issue will be addressed using renormalized MF theory in a future
work. An extension of our VMC study with simultaneous
inhomogeneous bond-order and singlet pair order parameters (as
required to treat such a problem) is difficult and also left for a
future work. Note also that low-energy {\it dynamic fluctuations}
of checkerboard (and SFP) characters could also exist within the d-wave RVB
state but this is beyond the scope of this paper.

The properties of the $BO/\mathcal{J}$ staggered flux wave-function are summarized
in Fig.~\ref{snapshots} showing the real and imaginary
parts of the measured hopping term $\langle c^+_{i}c_j \rangle$
between every nearest neighbor sites of our candidate
$BO/\mathcal{J}$ wave-function. We also present the exchange term
on each bonds of the lattice, and the local on-site charge
density. We find that the bond-order wave-function has both
(spin-spin) bond density wave and (small) charge density wave
components. Nonetheless, the charge modulations are very small
(the maximum charge deviation from the mean on-site charge is of
the order of $2\% $) , and the charge density is a little bit
larger in the center of the unit-cell. As expected, the
$SFP/\mathcal{J}$ has homogeneous hopping and exchange bonds
within the unit-cell. Therefore, we conclude that after projection
the modulated variational wave-function differs quantitatively
from the uniform one:
the $BO/\mathcal{J}$ staggered flux wave-function is quite inhomogeneous (although
with a very small charge modulation) leading to an increased magnetic
energy gain
while still preserving a competitive kinetic energy,
a characteristic of the homogeneous $SFP/\mathcal{J}$
wave-function.

\section{Conclusion}
\label{conclusion}

In conclusion, in this work we have investigated the $t{-}J{-}V$
model using both mean-field calculations as well as more involved
variational Monte-Carlo calculations. Both approaches have
provided strong evidence that bond-order wave-functions (of underlying
staggered flux character) are
stabilized at zero and finite Coulomb repulsion for doping close
to $1/8$. In particular, variational Monte-Carlo calculations show
that a bond modulation appears spontaneously
on top of the staggered flux phase. This is in agreement with the work of
Wang et al.~\cite{susceptibility} predicting an instability of staggered flux type.
We have also shown that
the modulated and homogeneous SFP, although nearby in parameter space,
are nevertheless separated from each other by a small energy barrier.
While both staggered flux wave-functions provide an optimal kinetic energy,
the bond-modulated one exhibits a small extra gain of the exchange energy.
On the other hand, a short range Coulomb repulsion
favors both staggered flux wave-function w.r.t the d-wave RVB superconductors and brings them
close in energy.

Finally, it would be interesting to study if the checkerboard
pattern could spontaneously appear in the vicinity of a vortex in
the mixed phase of the cuprates. Such an issue could be addressed
by studying the $t{-}J{-}V$ model on a square lattice extending
our variational scheme to include {\it simultaneously} nearest
neighbor pairing and bond modulated staggered currents. It is
expected that, while the pairing is suppressed in the vicinity of
the vortex, the checkerboard pattern might be variationally
stabilized in this region.

\acknowledgments We are grateful to Thierry Giamarchi and Andreas
L\"auchli for very useful discussions. This work was partially
supported by the Swiss National Fund and by MaNEP.


\begin{thebibliography}{}

\bibitem{anderson_hgtc_hubbard} P.W.~Anderson, Science {\bf 235},
1196 (1987).

\bibitem{Zhang-Rice} F.~C.~Zhang and T.~M.~Rice, {\bf 37}, 3759 (1988).
%abstract = {Effective Hamiltonian for the superconducting Cu oxides}

\bibitem{review_Dagotto} E.~Dagotto, Rev. of Mod. Phys. {\bf 66}, 763  (1994).

\bibitem{kotliar_fluxphases} G.~Kotliar, Phys.~Rev.~B {\bf 37},
3664 (1988). This state can also be written as a projected
\emph{s+id} spin liquid.

\bibitem{yokoyama_mcv_supra} H.~Yokoyama and H.~Shiba,
  J. Phys. Soc. Jpn. {\bf 57}, 2482 (1988).

\bibitem{gros_mcv_supra} C.~Gros, Phys.~Rev.~B {\bf 38}, R931
(1988); For recent estimations see e.g. A.~Paramekanti,
M.~Randeria and N.~Trivedi, Phys.~Rev.~Lett. {\bf 87}, 217002
(2001).

\bibitem{paramekanti_mcv} A.~Paramekanti, M.~Randeria and N.~Trivedi,
Phys.~Rev.~B {\bf 70}  , 054504 (2004).

\bibitem{RVB2} P.W.~Anderson, P.A.~Lee, M.~Randeria, T.M.~Rice, N.~Trivedi
and F.C.~Zhang, J Phys. Condens. Matter {\bf 16}, R755-R769
(2004).

\bibitem{half-flux} I.~Affleck and J.B.~Marston,
Phys.~Rev.~B. {\bf 37}, R3774 (1988); J.B.~Marston and I.~Affleck,
{\it ibid.} {\bf 39}, 11538 (1989).

\bibitem{staggered_flux}
D.A.~Ivanov, Phys.~Rev.~B. {\bf 70}, 104503 (2004) and references
therein; see also D.~Poilblanc and Y.~Hasegawa, Phys.~Rev.~B {\bf
41}, 6989 (1990).

\bibitem{PLee1}
P.A.~Lee,  N.~Nagaosa, T.K.~Ng and X.G.~Wen , Phys.~Rev.~B {\bf
57}, 6003 (1998).
%{SU(2) formulation of the t-J model: Application to underdoped cuprates}

\bibitem{PLee2}
X.G.~Wen and P.A.~Lee, Phys.~Rev.~Lett. {\bf 76}, 503 (1996).
%Theory of underdoped cuprates

\bibitem{PLee3}
M.U.~Ubbens and P.A.~Lee, Phys.~Rev.~B {\bf 46}, 8434 (1992).
%Flux phases in the t-J model

\bibitem{Hofstadter} D.R.~Hofstadter,  Phys.~Rev.~B. {\bf 14}, 2239 (1976).

\bibitem{Flux_phases_0} P.W.~Anderson, B.S.~Shastry and D.~Hristopulos,
Phys.~Rev.~B. {\bf 40}, 8939 (1989); D.~Poilblanc, {\it ibid.}
{\bf 40}, R7376 (1989); P.~Lederer, D.~Poilblanc and T.M.~Rice,
Phys.~Rev.~Lett. {\bf 63}, 1519 (1989); For related results using
slave boson MF techniques see e.g. F.~Nori, E.~Abrahams and
G.T.~Zimanyi, Phys.~Rev.~B. {\bf 41}, R7277 (1990).

\bibitem{Flux_phases_1} D.~Poilblanc, Y.~Hasegawa and T. M. Rice,
Phys.~Rev.~B. {\bf 41}, 1949 (1990).

\bibitem{STM-BSCO} M.~Vershinin, S.~Misra, S.~Ono, Y.~Abe,
Y.~Ando, and A.~Yazdani, Science {\bf 303}, 1995 (2004). Note
that the first observation was made around a vortex core in
J.E.~Hoffman, E.W.~Hudson, K.M.~Lang, V.~Madhavan, H.~Eisaki,
S.~Uchida, and J.C.~Davis, Science {\bf 295}, 466 (2002).

\bibitem{STM-BSCO-note} Note that the {\it energy-dependent}
spatial modulations of the tunneling conductance of optimally
doped BSCO can be understood in terms of elastic scattering of
quasiparticles. See J.E.~Hoffman, K.~McElroy, D.H.~Lee,
K.M.~Lang, H.~Eisaki, S.~Uchida, and J.C.~Davis Science {\bf 297}
1148 (2002).

\bibitem{STM_fourfold_structure} G.~Levy, M.~Kugler,
A.A.~Manuel, O.~Fischer and M.~Li
%Fourfold Structure of Vortex-Core States in Bi2Sr2CaCu2O8+
Phys.~Rev.~Lett. {\bf 95}, 257005 (2005).

\bibitem{STM_fourfold_charge_order}
A.~Hashimoto, N.~Momono, M.~Oda, and M.~Ido, cond-mat/0512496.
%STM/STS Study on 4a X 4a Electronic Charge Order and Inhomogeneous
%Pairing Gap in Superconducting Bi2Sr2CaCu2O8+d

\bibitem{STM2} T.~Hanaguri et al., Nature {\bf 430}, 1001 (2004).

\bibitem{note_distance} The Manhattan distance
of Ref.~\protect\onlinecite{Renormalized_MF_3} is used.

\bibitem{Gutzwiller} M.C.~Gutzwiller, Phys.~Rev.~Lett. {\bf 10}, 159 (1963);
D.~Vollhardt, Rev.~Mod.~Phys. {\bf 56}, 99 (1984).

\bibitem{Renormalized_MF} F.C.~Zhang, C.~Gros, T.M.~Rice and H.~Shiba,
Supercond.~Sci.~Technol.~{\bf 1}, 36 (1988).

\bibitem{Renormalized_MF_2} D.~Poilblanc, Phys.~Rev.~B. {\bf 41}, R4827 (1990);
note that, in this early treatment, uniform Gutzwiller parameters
were assumed although the bond variables were
allowed to vary spatially.

\bibitem{Renormalized_MF_3}
D.~Poilblanc, Phys.~Rev.~B. {\bf 72}, 060508(R) (2005).

\bibitem{Anderson_suggestion_4a}
Such a formulation should be appropriate as long as the deviations
of $\big<n_i\big>$ from the average density remain small. See
P.W.~Anderson, cond-mat/0406038; B.A.~Bernevig et al.,
cond-mat/0312573.

\bibitem{chemical_function} C.~Li, S.~Zhou and Z.~Wang,
cond-mat/0510596.

\bibitem{susceptibility} Z.~Wang, G.~Kotliar and X.F.~Wang, Phys.~Rev.~B.
{\bf 42}, R8690 (1990); note that the Fermi surface of the SFP is
made of four small elliptic-like \emph{pockets} centered around
$(\pm\pi/2,\pm\pi/2)$.

\bibitem{note_charge} Note that the bond modulation itself leads to
non-equivalent sites which, strictly speaking, should have
slightly different electron densities (although the $\epsilon_i$
might be constant).

\bibitem{note_AF} Note that, in this regime, antiferromagnetism is expected.
Such a competition is not considered here.

\bibitem{Vojta} Also found in $SU(2N)$ mean-field theory;
see M.~Vojta, Y.~Zhang and S.~Sachdev,
 Phys.~Rev.~B. {\bf 62}, 6721 (2000) and references therein.


\bibitem{Vojta2} Our $4\times 4$ solution bears some similarities
with those obtained within $SU(2N)$/$Sp(2N)$ mean field theories;
see M.~Vojta, Phys.~Rev.~B. {\bf 66}, 104505 (2002). Note however
that the large-N Sp(2N) scheme implies a superconducting state.

\bibitem{Jastrow_factor} R.~Jastrow, Phys.~Rev. {\bf 98}, 1479
(1955).

\end{thebibliography}
\end{document}